\documentclass[aps,prc,twocolumn,superscriptaddress,nofootinbib,floatfix,longbibliography]{revtex4-2}

\usepackage[T1]{fontenc}
\usepackage[utf8]{inputenc}
\usepackage{lmodern}
\usepackage{microtype}
\usepackage{graphicx}
\usepackage{booktabs}
\usepackage{amsmath,amssymb,bm,mathtools}
\usepackage{physics}
\usepackage{xcolor}
\usepackage{hyperref}
\usepackage{siunitx}
\usepackage{multirow}
\usepackage{placeins}
\usepackage{adjustbox}

\hypersetup{colorlinks=true,citecolor=blue,linkcolor=blue,urlcolor=blue}
\newcommand{\sNN}{\sqrt{s_{\mathrm{NN}}}}

\usepackage{lineno}

\begin{document}

\title{Predictions for neutron-tagged Coulomb dissociation in ultraperipheral Ru+Ru and Zr+Zr collisions}

\author{Huda Nasrulloh}
\author{Xinbai Li}
\author{Zebo Tang}
\author{Kaiyang Wang}
\email{kaiyangwang@ustc.edu.cn}
\author{Luobing Wang}
\author{Xin Wu}
\email{wuxinust@ustc.edu.cn}
\author{Wangmei Zha}

\affiliation{Department of Modern Physics, University of Science and Technology of China, Hefei, Anhui, China}
\date{\today}

\begin{abstract}

The photonuclear response reflects nuclear structure and the dynamics of excitation and decay, with neutron-emission channels retaining information that can be obscured in inclusive dissociation yields.
We calculate specified-side and mutual neutron-tagged Coulomb-dissociation cross sections in ultraperipheral $^{96}_{44}\mathrm{Ru}+{}^{96}_{44}\mathrm{Ru}$ and $^{96}_{40}\mathrm{Zr}+{}^{96}_{40}\mathrm{Zr}$ collisions at $\sqrt{s_{\mathrm{NN}}}=200~\mathrm{GeV}$ using an impact-parameter-dependent equivalent-photon framework with photonuclear inputs from TALYS calculations and ENDF evaluations.
Across the two input sets, the predicted Ru/Zr ratios span $1.54$--$1.71$ for exact $1n$ and $2.49$--$3.09$ for $1n1n$, while inclusive $Xn$ (at least one neutron) and $XnXn$ ratios remain near unity at $0.96$--$0.99$ and $1.05$--$1.10$, respectively.
The one-neutron ratios lie well above the corresponding flux-only $Z^2$ and $Z^4$ charge-scaling references, reflecting differences in the low-energy one-neutron response that are amplified by mutual tagging.
This topology hierarchy identifies neutron-tagged isobar ratios as sensitive probes of photonuclear response, revealing response differences hidden in inclusive dissociation.
These calculations also provide neutron-emission inputs for neutron-tagged vector-meson photoproduction in isobar collisions.
\end{abstract}

\maketitle

\section{Introduction}

Ultraperipheral collisions (UPCs) of relativistic heavy ions provide an especially clean environment for studying photonuclear reactions. At impact parameters larger than the region of significant hadronic overlap, short-range nuclear interactions are strongly suppressed, while the Lorentz-contracted electromagnetic fields generate intense fluxes of quasi-real photons that can dissociate the opposing nucleus~\cite{bertulani1988,baur2002,baltz2008,KleinSteinberg2020}. The emitted neutrons remain strongly collimated along the beam direction and can be detected in zero-degree calorimeters (ZDCs), providing experimental access to the neutron-emission channels of Coulomb dissociation~\cite{Adler2001,chiu2002,alice2023n,Baltz1998Luminosity,alice2012emd,alice2015rho}. In vector-meson photoproduction, either nucleus in a symmetric UPC can emit the photon, so the yield at a given nonzero rapidity combines contributions from two photon energies. Neutron classes sample different impact-parameter distributions and hence change the relative weights of these contributions, providing a way to resolve this two-way ambiguity~\cite{Guzey2014NeutronTagging}. CMS, ALICE, and STAR used the $0n0n$, single-side neutron-emission, and $XnXn$ classes to extract the coherent $J/\psi$ photonuclear cross section as a function of the photon--nucleon center-of-mass energy $W_{\gamma N}$ through a combined fit of the corresponding linear relations, enabling studies of nuclear gluon distributions and possible gluon saturation~\cite{CMS2023JpsiNeutron,ALICE2023JpsiNeutron,STAR2024JpsiNeutron}. The coefficients in these relations are photon fluxes folded with the impact-parameter-dependent neutron-emission probabilities, making Coulomb-excitation and neutron-emission calculations essential inputs to this method. Neutron tagging has also been studied as a means of controlling the impact-parameter distribution in electron--ion photoproduction~\cite{wu2025impact}, while angular correlations in two-photon dilepton production provide a complementary probe of heavy-ion collision geometry~\cite{Luo2026CollisionGeometry}.

Photonuclear excitation is a fundamental probe of nuclear structure and collective dynamics. In the giant dipole resonance (GDR) region, photon absorption is dominated by electric-dipole excitation, in which the proton and neutron distributions oscillate against one another~\cite{Bracco2019Dipole}. In this isovector mode, the resonance energy reflects the restoring force for relative proton--neutron motion, while the energy distribution of the strength is sensitive to nuclear shape~\cite{Junghans2008GDRShape,Bracco2019Dipole}. Studies of the complete dipole response and its polarizability have, in turn, provided constraints on neutron skins and the symmetry energy---the energy cost of neutron--proton imbalance in nuclear matter~\cite{Tamii2011DipoleResponse,Piekarewicz2012DipoleSkin}. Photon absorption above the neutron-separation threshold can be followed by neutron emission. The photoneutron cross section for a given neutron multiplicity reflects both the energy-dependent absorption strength and the subsequent decay of the excited nucleus~\cite{Goriely2020E1Moments,RIPL3,koning2012}. An exact $1n$ tag selects events with precisely one emitted neutron and emphasizes the GDR-region one-neutron response. An inclusive $Xn$ tag accepts all events with at least one neutron, including higher-multiplicity channels that become accessible as the excitation energy increases~\cite{Baltz1998Luminosity,Pshenichnov2001,Pshenichnov2011,klusek2013emd}. The distinction between exact and inclusive neutron emission also extends to mutual Coulomb dissociation, in which both nuclei are electromagnetically excited in the same UPC event~\cite{Baltz1998Luminosity,Baltz2002,Klein2017}. Resolving these channels therefore provides a joint test of photonuclear excitation and neutron decay, allowing nuclear-response differences to emerge even when inclusive yields are similar.

Ru+Ru and Zr+Zr collisions at RHIC provide a controlled isobar setting for this comparison. They involve $^{96}$Ru ($Z=44$, $N=52$) and $^{96}$Zr ($Z=40$, $N=56$), which have the same mass number and comparable nuclear size but different charge and neutron content. The larger Ru charge increases the equivalent-photon flux, while differences in neutron-separation energies and channel-resolved photoneutron cross sections modify the nuclear response~\cite{Wang2021AME2020}. These flux and response effects can reinforce or compensate one another, depending on the neutron channel selected. The central question is whether neutron selection can expose differences in nuclear excitation and decay that are obscured in inclusive isobar yields, and how mutual tagging enhances this sensitivity. Established descriptions of electromagnetic dissociation include RELDIS~\cite{Pshenichnov2001,Pshenichnov2011} and the forward-neutron generator nOOn~\cite{nOOn2020}. To our knowledge, dedicated predictions for specified-side and mutual neutron-tagged Coulomb dissociation in $^{96}\mathrm{Ru}+{}^{96}\mathrm{Ru}$ and $^{96}\mathrm{Zr}+{}^{96}\mathrm{Zr}$ collisions at $\sqrt{s_{\mathrm{NN}}}=200~\mathrm{GeV}$ have not been available in the published literature.

In this work, we test how neutron selection changes the balance between the photon field and nuclear response by calculating specified-side and mutual Coulomb-dissociation cross sections for Ru+Ru and Zr+Zr at $\sqrt{s_{\mathrm{NN}}}=200~\mathrm{GeV}$ within an impact-parameter-dependent equivalent-photon approximation (EPA). We use TALYS and ENDF photonuclear inputs to examine the input dependence of the cross sections and the persistence of the neutron-topology hierarchy. The calculation also provides the neutron-emission probabilities needed to construct neutron-class-dependent photon fluxes for isobar vector-meson photoproduction. Section~\ref{sec:model} introduces the photonuclear inputs and develops the impact-parameter formulation of specified-side and mutual neutron-tagged cross sections. Section~\ref{sec:results} compares the calculation with published Au+Au, Pb+Pb, and U+U results and analyzes the Ru/Zr cross sections and topology-dependent ratios. Section~\ref{sec:conclusion} summarizes the main conclusions.

\section{Photonuclear inputs and calculation framework}
\label{sec:model}

The neutron-tagged cross sections are obtained by folding photonuclear reaction cross sections with the equivalent-photon density and integrating the resulting event probabilities over impact parameter with the UPC survival factor.
At photon energy $\omega$ in the rest frame of nucleus $A$, $\sigma_{\gamma A\rightarrow Nn}(\omega)$ denotes the cross section for a single-photon reaction emitting exactly $N$ neutrons. The inclusive $Xn$ channel includes all neutron-producing reactions, so its cross section satisfies
\[
\sigma_{\gamma A\rightarrow Xn}(\omega)
=\sum_{N\geq1}\sigma_{\gamma A\rightarrow Nn}(\omega).
\]
Each reaction is counted once, irrespective of the emitted neutron multiplicity. The calculation requires the exact-$1n$ and inclusive-$Xn$ inputs.

For the exact-$1n$ channel, we use TALYS photonuclear-reaction calculations or evaluated neutron-emission data from the ENDF library~\cite{koning2012,iaea-endf}. These are two sources of photonuclear input to the same Coulomb-dissociation framework. The exact-$1n$ response is dominated by GDR-region dipole excitation followed by one-neutron decay. For Ru and Zr, we compare these inputs with the RIPL-3 Lorentzian GDR parameterization,
\begin{equation}
\label{eq:gdr_lorentzian}
\sigma_{\rm GDR}(E_\gamma)=
\sum_m \sigma_{0m}\,
\frac{\Gamma_m^2E_\gamma^2}
{(E_\gamma^2-E_{0m}^2)^2+\Gamma_m^2E_\gamma^2},
\end{equation}
where $E_\gamma\equiv\omega$, and $E_{0m}$, $\Gamma_m$, and $\sigma_{0m}$ denote the centroid energy, width, and peak cross section of component $m$. The sum allows more than one GDR component~\cite{RIPL3}. Nuclear deformation can split the collective dipole strength among oscillations along different axes, while coupling to more complex nuclear configurations contributes to its broadening~\cite{Junghans2008GDRShape}. This reference therefore characterizes the gross dipole-absorption response rather than an exact neutron-emission channel. After excitation, neutron emission competes with other decay modes. The neutron-separation energies and level densities of accessible states determine how the excitation strength is distributed among the available neutron final states~\cite{RIPL3,koning2012}. The TALYS and ENDF exact-$1n$ inputs therefore incorporate the combined effects of photon absorption and subsequent decay selection.

The neutron thresholds help interpret the contrast between the two isobars. The GDR maximum lies below the two-neutron separation threshold in $^{96}$Ru but near this threshold in $^{96}$Zr~\cite{Wang2021AME2020}, leaving a broader GDR-energy interval in Ru in which neutron emission is restricted to the exact-$1n$ channel.

The inclusive $Xn$ input includes neutron-emission strength in the GDR region together with contributions at higher photon energies. Its higher-energy continuation is based on empirical photoabsorption systematics~\cite{Caldwell1973,Armstrong1972,Michalowski1977,Carlos1984,Lepretre1978,Veyssiere1970}, following the phenomenology used in established electromagnetic-dissociation calculations, including RELDIS~\cite{Pshenichnov2001,Pshenichnov2011} and nOOn~\cite{nOOn2020}. This continuation covers the quasi-deuteron, resonance, and continuum regions, including the high-energy photoabsorption regimes, with nuclear shadowing included at high photon energies.

\begin{figure*}[!tbp]
\centering
\includegraphics[width=0.95\textwidth]{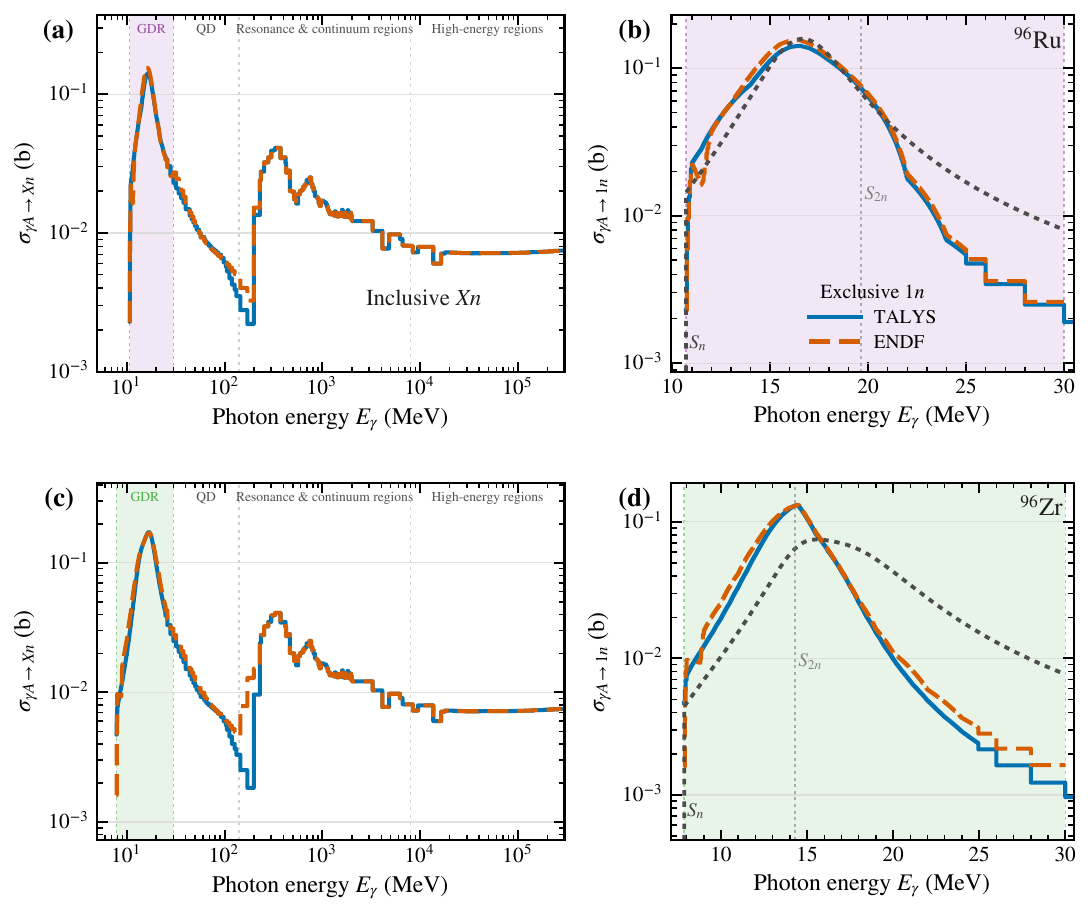}
\caption{(Color online) Photonuclear inputs for $^{96}$Ru (top) and $^{96}$Zr (bottom). Panels (a) and (c) show the inclusive
$\gamma A\rightarrow Xn$ cross sections over a broad photon-energy range, while panels (b) and (d) show the exclusive
$\gamma A\rightarrow1n$ response in the GDR region. Blue solid and orange dashed curves denote the TALYS- and ENDF-based inputs, respectively. The dotted curves in panels (b) and (d) show the RIPL-3 standard-Lorentzian GDR reference~\cite{RIPL3}. Purple and green shading identify the Ru and Zr GDR regions, respectively, and the vertical $S_n$ and $S_{2n}$ markers indicate the one- and two-neutron separation thresholds.}
\label{fig:isobar_xn_inputs}
\end{figure*}

Figure~\ref{fig:isobar_xn_inputs} collects the Ru and Zr inputs. Panels (b) and (d) resolve the GDR-region exact-$1n$ response, while panels (a) and (c) display the inclusive $Xn$ cross sections over the broader photon-energy range. These two inputs enter the same EPA folding below, allowing the neutron selection to retain different parts of the nuclear response.

To convert these photonuclear inputs into event probabilities, we fold them with the electromagnetic field of the opposing ion. For an emitting nucleus of charge $Z$, the EPA photon density per unit photon energy and transverse area, $n(\omega,b)\equiv d^3N_\gamma/(d\omega\,d^2b)$, is given in the ultrarelativistic point-charge approximation by
\begin{equation}
\label{eq:flux_b}
n(\omega,b)=
\frac{Z^2\alpha}{\pi^2\omega b^2}\,
\xi^2\!\left[
K_1^2(\xi)+\gamma^{-2}K_0^2(\xi)
\right],
\hspace{0.2cm}
\xi=\frac{\omega b}{\gamma\hbar c}.
\end{equation}
Here $\gamma$ is the Lorentz factor of one ion in the rest frame of the other, and $K_0$ and $K_1$ are modified Bessel functions. In symmetric collider systems, $\gamma=\gamma_{\rm rel}=2\gamma_{\rm beam}^2-1$, where $\gamma_{\rm beam}=\sqrt{s_{\rm NN}}/(2m_N)$.

For excitation of nucleus $A$, the photon flux is generated by the opposing nucleus. At fixed impact parameter, the mean number of single-photon reactions in neutron channel $i$ is
\begin{equation}
\label{eq:mu_tag}
\mu_i(b)=\int d\omega\,n(\omega,b)\,
\sigma_{\gamma A\rightarrow i}(\omega),
\qquad i\in\{1n,Xn\}.
\end{equation}
At fixed $b$, the number of neutron-producing excitations is taken to follow a Poisson distribution~\cite{Baltz1996,Pshenichnov2001,Pshenichnov2011,klusek2013emd,wu2025impact}. The probability of no such excitation is
\begin{equation}
\label{eq:p0}
 P_{0n}(b)=\exp[-\mu_{Xn}(b)],
\end{equation}
and the inclusive probability for at least one emitted neutron is
\begin{equation}
\label{eq:pxn}
 P_{Xn}(b)=1-\exp[-\mu_{Xn}(b)].
\end{equation}
For the exact-$1n$ tag, the probability is
\begin{equation}
\label{eq:p1n}
P_{1n}(b)=\mu_{1n}(b)\exp[-\mu_{Xn}(b)].
\end{equation}
An exact-$1n$ event thus requires one excitation in the $1n$ channel and no additional neutron-producing excitation.

Hadronic overlap is excluded through the common event-level UPC survival probability,
\begin{equation}
\label{eq:survival}
\begin{aligned}
S_{\rm UPC}(b)&=\exp[-P_H(b)],\\
P_H(b)&=
\sigma_{NN}^{\rm inel}
\int d^2s\,
T_A({\bf s})T_B({\bf b}-{\bf s}) .
\end{aligned}
\end{equation}
Here $P_H(b)$ is the mean number of inelastic nucleon--nucleon collisions in the optical Glauber approximation~\cite{Miller2007Glauber}. The thickness functions are $T_A(\mathbf{s})=\int dz\,\rho_A(\mathbf{s},z)$, with $\int d^2s\,T_A(\mathbf{s})=A$ and similarly for $B$, using Woods--Saxon density profiles~\cite{WoodsSaxon1954,DeVries1987}. The inelastic nucleon--nucleon cross section $\sigma_{NN}^{\rm inel}$ is chosen for the collision energy under consideration~\cite{Wang2025DiHadron}.

The specified-side neutron-tag cross section for nucleus $A$ is
\begin{equation}
\label{eq:sigma_tag}
\sigma_i^{(A)}
=
2\pi\int b\,db\,
S_{\rm UPC}(b)\,
P_i^{(A)}(b),
\end{equation}
where $P_i^{(A)}(b)$ is the probability that nucleus $A$ satisfies neutron class $i$, while no condition is imposed on the neutron-emission state of the opposite nucleus. For mutual Coulomb dissociation into neutron classes $i$ and $j$,
\begin{equation}
\label{eq:sigma_mutual}
\sigma_{ij}^{(A,B)}
=
2\pi\int b\,db\,
S_{\rm UPC}(b)\,
P_i^{(A)}(b)P_j^{(B)}(b).
\end{equation}
The neutron-emission probabilities for the two nuclei factorize at the same $b$, and the common hadronic-survival factor is applied once per event. The $1nXn$ observable is ordered, with the $1n$ condition applied to one specified nucleus and the $Xn$ condition applied to the opposite nucleus. Thus, $1n1n$ events are included in the $1nXn$ observable.

\section{Results and discussion}
\label{sec:results}

\begin{figure*}[!t]
\centering
\includegraphics[width=0.95\textwidth]{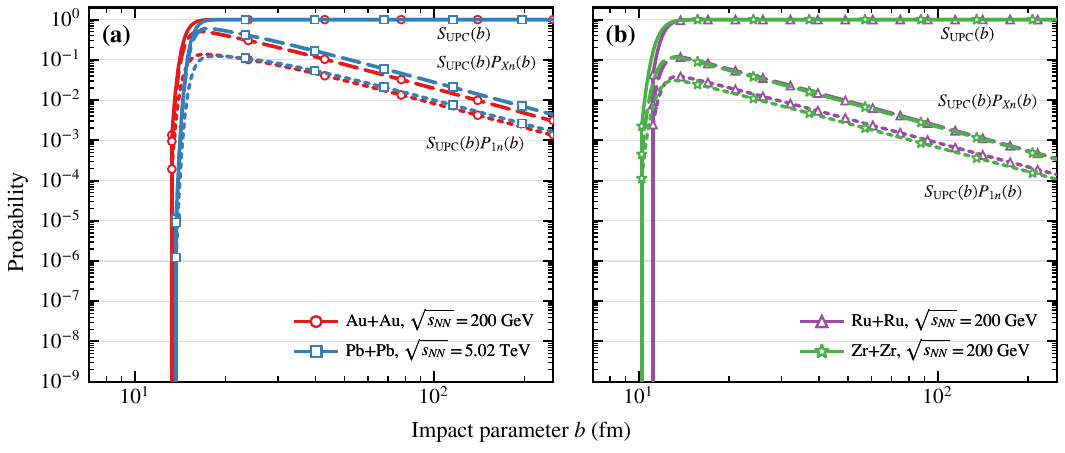}
\caption{(Color online) Impact-parameter dependence of the survival-weighted neutron probabilities for Au+Au and Pb+Pb in panel (a) and Ru+Ru and Zr+Zr in panel (b). Red circles, blue squares, purple triangles, and green stars identify Au, Pb, Ru, and Zr. Solid, dashed, and dotted curves denote $S_{\mathrm{UPC}}(b)$, $S_{\mathrm{UPC}}(b)P_{Xn}(b)$, and $S_{\mathrm{UPC}}(b)P_{1n}(b)$. The turn-on at the nuclear edge follows the hadronic-survival factor, while the large-$b$ decrease of the neutron-emission curves follows the equivalent-photon density.}
\label{fig:probability_vs_b}
\end{figure*}

The neutron-tagged Coulomb-dissociation cross section combines the equivalent-photon field, the channel-resolved photonuclear response, and the UPC survival probability. Toward smaller $b$, the stronger photon field increases electromagnetic excitation, while hadronic overlap suppresses the UPC contribution. At large $b$, $S_{\rm UPC}(b)$ approaches unity, but the photon density decreases approximately as $1/b^2$ in the small-$\xi$ limit.
Figure~\ref{fig:probability_vs_b} illustrates this competition for Au+Au and Pb+Pb in panel (a) and for the isobars in panel (b). The solid curves show the survival probability, while the dashed and dotted curves show the survival-weighted $Xn$ and $1n$ probabilities, respectively. In panel (b), the purple Ru and green Zr curves are clearly separated for $1n$ but much closer for $Xn$. This channel-dependent contrast provides a first indication that neutron selection changes which differences in the photonuclear response are retained in the observable cross sections.

The integrated cross sections are benchmarked in Table~\ref{tab:benchmark_compact} against published Au+Au results at RHIC~\cite{Baltz1996,Baltz1998Luminosity,Pshenichnov2001}, Pb+Pb results at the LHC~\cite{alice2023n}, and U+U results at RHIC~\cite{fischer2014uranium}. The inclusive $Xn$ and $XnXn$ cross sections agree with the reference results at the few-percent level, supporting the common survival-weighted EPA integration. The exact-$1n$ predictions show a larger variation between photonuclear inputs, reflecting their sensitivity to the detailed low-energy one-neutron response.

\begin{table}[!tbp]
\centering
\caption{Benchmark comparison for specified-side neutron tags ($1n$
and $Xn$) and mutual Coulomb dissociation ($XnXn$).  The $1n$ ranges show the
spread between low-energy photonuclear inputs; they are not statistical
uncertainties.  The reference column contains representative published
calculations or measurements at the energies reported in the cited publications. All
cross sections are in barns.}
\label{tab:benchmark_compact}
\footnotesize
\setlength{\tabcolsep}{3.5pt}
\begin{ruledtabular}
\begin{tabular}{lcc}
System/channel & This work & Reference \\
\hline
Au $1n$        & $48.51$--$55.20$  & $49.00$--$51.00$~\cite{Baltz1998Luminosity} \\
Au $Xn$        & $94.91$           & $95.00$--$95.10$~\cite{Baltz1996,Pshenichnov2001} \\
Au $XnXn$      & $3.93$            & $3.80$--$3.90$~\cite{Baltz1998Luminosity,Pshenichnov2001} \\
Pb $1n$        & $100.55$--$103.71$ & $103.70$--$108.40$~\cite{alice2023n} \\
Pb $Xn$        & $215.72$          & $204.60$--$220.00$~\cite{Baltz1996,alice2023n,fischer2014uranium} \\
Pb $XnXn$      & $6.58$            & $6.20$~\cite{Pshenichnov2001,fischer2014uranium} \\
U $Xn$         & $154.18$          & $150.10$~\cite{fischer2014uranium} \\
U $XnXn$       & $8.08$            & $7.59$~\cite{fischer2014uranium} \\
\end{tabular}
\end{ruledtabular}
\end{table}

\begin{table*}[!t]
\centering
\caption{Calculated neutron-tagged Coulomb-dissociation cross sections for Ru+Ru and Zr+Zr at $\sNN=200~\mathrm{GeV}$. The specified-side
$\sigma_{1n}$ and $\sigma_{Xn}$ cross sections are given in barns, while the mutual-dissociation channels are given in millibarns. Here $Xn$ denotes emission of at least one neutron. The $1nXn$ entry follows the ordered convention defined in Sec.~\ref{sec:model}. The TALYS--ENDF spread reflects photonuclear-input dependence and does not represent the full theoretical uncertainty. The corresponding Ru/Zr ratios are shown in Fig.~\ref{fig:ratio}.}
\label{tab:isobar_summary}
\begin{ruledtabular}
\begin{tabular}{lcccccc}
Observable
& \multicolumn{2}{c}{Ru+Ru}
& \multicolumn{2}{c}{Zr+Zr}
& \multicolumn{2}{c}{Ru/Zr} \\
\cline{2-3}\cline{4-5}\cline{6-7}
& TALYS & ENDF & TALYS & ENDF & TALYS & ENDF \\
\hline
$\sigma_{1n}$ [b]
& 5.73 & 5.58
& 3.35 & 3.62
& 1.71 & 1.54 \\

$\sigma_{Xn}$ [b]
& 10.89 & 11.30
& 10.99 & 11.71
& 0.99 & 0.96 \\

$\sigma_{1n1n}$ [mb]
& 17.90 & 16.62
& 5.79 & 6.66
& 3.09 & 2.49 \\

$\sigma_{1nXn}$ [mb]
& 47.76 & 47.30
& 25.92 & 29.27
& 1.84 & 1.62 \\

$\sigma_{XnXn}$ [mb]
& 127.73 & 134.96
& 116.25 & 128.88
& 1.10 & 1.05 \\
\end{tabular}
\end{ruledtabular}
\end{table*}

\begin{figure*}[!tbp]
\centering
\includegraphics[width=0.95\textwidth]
{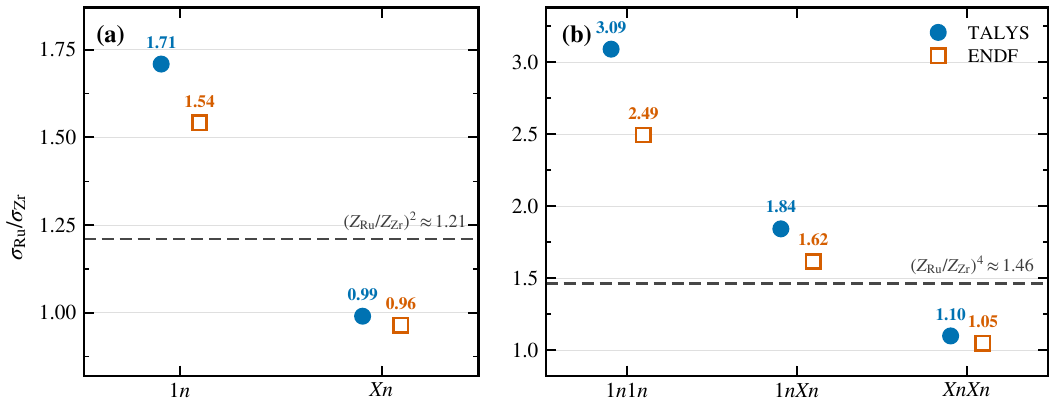}
\caption{(Color online) Ru/Zr cross-section ratios for specified-side neutron tags in panel (a) and mutual Coulomb dissociation in panel (b). Blue filled circles and orange open squares denote results obtained with the TALYS and ENDF inputs. Black dashed lines show the flux-only references of $1.21$ for specified-side tags and $1.46$ for mutual dissociation. Exact $1n$ and $1n1n$ tags lie above these references, the ordered $1nXn$ tag is intermediate, and the inclusive $Xn$ and $XnXn$ ratios remain close to unity.}
\label{fig:ratio}
\end{figure*}

The Ru+Ru and Zr+Zr cross sections at $\sNN=200~\mathrm{GeV}$ are collected in Table~\ref{tab:isobar_summary}, and the corresponding ratios $R_i\equiv\sigma_i^{\rm Ru+Ru}/\sigma_i^{\rm Zr+Zr}$ for each neutron topology are shown in Fig.~\ref{fig:ratio}. Panel (a) compares specified-side $1n$ and $Xn$ tags, while panel (b) compares the mutual neutron topologies; blue filled circles and orange open squares denote TALYS and ENDF, respectively. The main result is a pronounced topology dependence: exact $1n$ and $1n1n$ tags show a large Ru--Zr separation, inclusive $Xn$ and $XnXn$ ratios remain near unity, and the mixed $1nXn$ ratio is intermediate.

For orientation, the black dashed lines show the simple flux-only estimates $(Z_{\rm Ru}/Z_{\rm Zr})^2=(44/40)^2\simeq1.21$ for a specified-side tag and $(Z_{\rm Ru}/Z_{\rm Zr})^4\simeq1.46$ for mutual dissociation. These estimates follow in the weak-excitation limit when differences in photonuclear response and geometry are neglected, and serve as reference points rather than complete predictions. The contrasting positions of the exact and inclusive tags relative to these common references highlight the role of the channel-resolved nuclear response.

The specified-side $1n$ tag gives Ru/Zr ratios of $1.71$ for TALYS and $1.54$ for ENDF, well above the flux-only reference. This enhancement reflects the stronger GDR-region exact-$1n$ response of Ru relative to Zr in Fig.~\ref{fig:isobar_xn_inputs}. Through Eq.~\eqref{eq:mu_tag}, the tag samples the energy-dependent partial $1n$ strength with the photon-spectrum weight, which favors lower photon energies in the small-$\xi$ limit. The neutron-threshold ordering discussed in Sec.~\ref{sec:model} helps explain this response contrast. The exact-$1n$ yield combines this channel contrast with the exclusion of additional neutron-producing excitations through Eq.~\eqref{eq:p1n}.
The inclusive $Xn$ tag shows the complementary behavior, with ratios of $0.99$ for TALYS and $0.96$ for ENDF. It sums all neutron-producing channels and samples a broader excitation range, so strength outside the exact-$1n$ branch also contributes to the inclusive input. For the adopted inputs, the stronger inclusive Zr response compensates for the larger Ru photon flux in the folded cross sections. The near-unity ratios therefore reflect a balance between the photon field and the inclusive nuclear response, rather than identical Ru and Zr photonuclear dynamics.
The channel information can be made explicit through the exact-$1n$ fraction within the specified-side $Xn$ yield. From Table~\ref{tab:isobar_summary}, $\sigma_{1n}/\sigma_{Xn}=0.49$--$0.53$ for Ru, compared with approximately $0.31$ for Zr. These UPC yield fractions include the photon-energy folding and event probabilities, rather than representing decay branching fractions at a fixed excitation energy. Thus, nearly equal inclusive yields conceal an enhancement of the exact-$1n$ fraction in Ru by a factor of $1.60$--$1.73$.

Applying neutron conditions to both nuclei amplifies the channel contrast in Fig.~\ref{fig:ratio}(b). At a common impact parameter, the mutual probability contains the product of the two specified-side probabilities. Selecting exact $1n$ on both sides therefore carries the Ru--Zr contrast into both factors, giving $1n1n$ ratios of $3.09$ for TALYS and $2.49$ for ENDF. The mixed $1nXn$ ratios, $1.84$ and $1.62$, are intermediate because only one side is restricted to the exact one-neutron channel. The fully inclusive $XnXn$ ratios, $1.10$ and $1.05$, remain near unity as the inclusive-response compensation acts on both sides.
Most of the large $1n1n$ contrast follows from the product amplification of the specified-side difference. The mutual ratios are not exactly the products of the integrated specified-side ratios, however, because the two probabilities are multiplied at the same $b$ before the survival-weighted integration. Mutual tagging gives greater weight to the smaller-$b$ UPC region with larger excitation probabilities, thereby reweighting the impact-parameter-dependent Ru--Zr contrast. Numerically, $R_{1n1n}/R_{1n}^{2}=1.06$ for TALYS and $1.05$ for ENDF, while $R_{XnXn}/R_{Xn}^{2}\simeq1.12$--$1.13$. These residual deviations reflect the different impact-parameter weighting in addition to the channel contrast selected on both sides.

Although the numerical ratios differ between TALYS and ENDF, both give the hierarchy $R_{1n1n}>R_{1nXn}>R_{XnXn}$ together with $R_{1n}>R_{Xn}$. This common ordering links the specified-side channel selection to its amplification in mutual dissociation. The hierarchy makes neutron-tagged Ru/Zr ratios a channel-resolved probe of photonuclear response, exposing channel-dependent response differences hidden in inclusive dissociation.

\FloatBarrier
\section{Conclusion}
\label{sec:conclusion}

We have calculated specified-side and mutual neutron-tagged Coulomb dissociation in $^{96}$Ru+$^{96}$Ru and $^{96}$Zr+$^{96}$Zr UPCs at $\sNN=200~\mathrm{GeV}$ using an impact-parameter-dependent EPA framework. Comparisons with published Au+Au, Pb+Pb, and U+U results support the calculation at the few-percent level for inclusive channels.
Both the TALYS and ENDF inputs yield a pronounced neutron-topology hierarchy. Exact $1n$ tags show a substantial Ru--Zr separation, which is amplified in mutual $1n1n$ dissociation, while the inclusive $Xn$ and $XnXn$ ratios remain near unity. The mixed $1nXn$ ratio lies between the exact and inclusive mutual ratios.
The hierarchy reflects how neutron selection retains different parts of the photonuclear response. Exact-$1n$ tags preserve the Ru--Zr contrast in the GDR-region one-neutron response. The inclusive ratios instead reflect compensation between the larger Ru photon flux and the stronger inclusive Zr response. Mutual tagging applies the channel selection to both nuclei at a common impact parameter, amplifying the response contrast with a different weighting of the UPC geometry.

The predicted hierarchy shows that similar inclusive dissociation cross sections can conceal substantial differences in nuclear excitation and decay. Measurements of specified-side and mutual neutron topologies would provide joint, channel-resolved tests of photonuclear-excitation and neutron-decay models, complementary to conventional photoabsorption measurements. Neutron-tagged isobar UPCs thus provide a differential probe of photonuclear response, making structure-dependent excitation and decay visible beyond inclusive dissociation and simple charge scaling. The calculated impact-parameter-dependent neutron-emission probabilities also supply the inputs needed to construct neutron-class-dependent photon fluxes for Ru+Ru and Zr+Zr photoproduction. These inputs are required to extend the neutron-tagging method for resolving the two-way photon-energy ambiguity to the isobar systems.

\begin{acknowledgments}
We thank Prof.~Yifei Niu for useful discussions. This work was supported in part by the National Natural Science Foundation of China under Grant Nos.~12605233, 12422510, and 12361141827, and the National Key Research and Development Program of China under Grant No.~2022YFA1604900. W.~Zha also acknowledges support from the Anhui Provincial Natural Science Foundation under Grant No.~2508085JX002, the Youth Innovation Promotion Association of the Chinese Academy of Sciences, and the Chinese Academy of Sciences under Grant No.~YSBR088.
\end{acknowledgments}

\clearpage
\bibliography{upc_1n_refs}
\end{document}